\documentstyle[prl,aps,multicol,psfig]{revtex}

\def\addcontentsline#1#2#3{\relax}
\begin{document}
\draft
\title{Coulomb drag between one-dimensional conductors }
\author{V.V. Ponomarenko$^*$ and D.V. Averin}
\address{Department of Physics and Astronomy, SUNY Stony Brook,
Stony Brook, NY 11794 }
\date{\today}
\maketitle
\begin{abstract}
We have analyzed Coulomb drag between currents of interacting
electrons in two parallel one-dimensional conductors of finite
length $L$ attached to external reservoirs. For strong coupling,
the relative fluctuations of electron density in the conductors
acquire energy gap $M$. At energies larger than $\Gamma = const
\times  v_- \exp (-LM/v_-)/L + \Gamma_{+}$, where $\Gamma_{+}$
is the impurity scattering rate, and for $L>v_-/M$, where $v_-$
is the fluctuation velocity, the gap leads to an ``ideal'' drag
with almost equal currents in the conductors. At low energies
the drag is suppressed by coherent instanton tunneling, and
the zero-temperature transconductance vanishes, indicating the
Fermi liquid behavior.

\end{abstract}


%
\multicols{2}

Current drag between two parallel conductors provides important
information about electron correlations both between the conductors
and inside them. The dominant drag mechanism is the momentum
transfer between electrons in the two conductors by Coulomb
scattering, and in the case of Fermi liquids, the usual phase-space
arguments applied to this scattering show that the linear drag
trans-conductance (and trans-resistance) should vanish as $T^2$ when 
temperature $T$ goes to zero. Vanishing zero-temperature
transconductance can be viewed then as a manifestation of the
Fermi-liquid behavior. In this work, we study the low-energy behavior 
of the current drag between one-dimensional conductors of interacting 
electrons
\cite{d1,d2,d3}, where the near-ideal equality of currents in the
two conductors was predicted \cite{d3} in the strong-coupling
regime. It is shown below, however, that despite this ``ideal'' drag
at large energies, the reservoirs and impurity scattering suppress
the drag at low energies making the linear
zero-temperature trans-conductance zero. Experimentally, this behavior 
of the current drag can be studied in coupled and individually contacted 
one-dimensional AlGaAs/GaAs heterostructures \cite{ex}, where the drag has 
been found recently \cite{debray,tarucha}. 

The set-up we consider consists of two identical parallel 1D
conductors of finite length $L$ attached to external reservoirs.
Our description of the current drag is based on the sin-Gordon model. 
In the language of this model, when the conductors are infinitely long,  
strong coupling between the currents in the two conductors
is equivalent to opening of the energy gap $M$ in the spectrum of
the relative electron density fluctuations. This requires an interacting repulsive
Tomonaga-Luttinger liquid (TLL) in the individual conductors.
For conductors of finite length $L$, the strong coupling is
expected if $M>T_L \equiv v_-/L$, where $v_-$ is velocity of the
relative electron density fluctuations. Here and below $e=\hbar=
1$. Deviations between the
currents in the two conductors in this regime are due to
tunneling of instantons of energy $M$. To describe this tunneling
we derive below a low-energy model using the Schmid's
duality transformation \cite{schmid}, and solve it
by fermionization. The main prediction that follows from the
solution is the existence of a new low-energy regime of coherent
multi-instanton tunneling that leads to a complete suppression
of the Coulomb drag at small temperatures and voltage differences
$V$ across the two conductors. Inclusion of 
a weak one-electron impurity scattering with small
rate $\Gamma_+$ suppresses
the trans-conductance further, increasing the energy range $\Gamma$ of the coherent
multi-instanton tunneling as $\Gamma = \mbox{const} \times \exp
(-LM/v_-)v_-/L +\Gamma_+$. 

Transport through a one-channel wire confined between two reservoirs
of spinless electrons and close to a screening gate
can be modeled  \cite{1d1,1d2} by a 1D system
of electrons whose interaction is local
and switched off outside
the finite length of the wire. Haldane's bosonization \cite{hald}
applied to each of the two wires relates their 1D electron densities
$\rho_b$ ($b=1,2$) to the appropriate bosonic fields $\phi_b$ as
$\rho_{b}(x,t)=(2 k_{bF}+\partial_x \phi_{b}(x,t))\sum \exp \{i
n(k_{bF}x+\phi_b(x,t)/2)\}/2 \pi$. Here summation runs over even $n$,
and $k_{bF}$, $b=1,2$, are the Fermi momenta of the two wires.
Evolution of each bosonic field $\phi_b(x,t)$ is described with
the Lagrangian of an inhomogeneous TLL
\[ {\cal L}_{b} = \int dx \frac{1}{2g_b(x)}\{ \left({{\partial_t
\phi_b(t,x) }\over {\sqrt{4 \pi}}} \right)^2 -
\left({{\partial_x \phi_b(t,x) } \over {\sqrt{4 \pi}}} \right)^2
\} \, ,
\]
where the coordinate $x$ was scaled by inverse velocity. 
The inhomogeneous interaction constant
of the TLL model, $g(x)= 1+(g-1) \varphi(x)$,
$\varphi(x)= \theta (x) \theta (L-x)$, corresponds to non-interacting
electrons
outside the wire,  $g(x)= 1$, and takes on some interacting value
$g(x)=g<1$ inside it.  The fields $\phi_\pm \equiv
(\phi_1 \pm \phi_2)/\sqrt{2} $
related to the fluctuations of the total and relative electron
densities in the wires have the Lagrangians ${\cal L}_\pm$ of
the same form, and the total Lagrangian
is ${\cal L}= \sum_\pm {\cal L}_\pm = \sum_{b=1,2} {\cal L}_b $.
Inter-wire interaction adds another part to the total Lagrangian,
$
{\cal L}_{int}= -\int^L_0 dx \ dy U(x-y) \rho_1(x) \rho_2(y)
$,
where $U(x-y)$ is short-ranged on the scale of the conductors' length $L$.
Substituting bosonic expression
for the densities into ${\cal L}_{int}$ we write the total
Lagrangian as
\endmulticols
\vspace{-6mm}\noindent\underline{\hspace{87mm}}
\begin{equation}
{\cal L} = \int dx \bigl[ \sum_\pm
\{ \left({{\partial_t \phi_\pm(t,x) } \over {g(x) \sqrt{8 \pi}}}
\right)^2 - {\pi\pm \varphi(x) U_0 g \over \pi}
\left({{\partial_x \phi_\pm(t,x) } \over {g(x)\sqrt{8 \pi}}} \right)^2
\} - {E_F^2 U_1 \over 2 \pi ^2} \varphi(x) \cos(2 \Delta k_F  x +
\sqrt{2} \phi_-(t,x)) \bigr] \, ,
\label{L} \end{equation}
\noindent\hspace{92mm}\underline{\hspace{87mm}}\vspace{-3mm}
\multicols{2}
\noindent
where $U_{0,1}/(2\pi)$ are constants that for weak potential $U(x)$
are equal to the amplitudes of the forward and backward scattering, 
and $\Delta k_F=k_{1F}-k_{2F}$. Equation
(\ref{L}) means that the forward scattering modifies the constants
$g_\pm$ of the TLL interactions and velocities of the modes:
$g_\pm= g/\sqrt{1\pm g U_0/\pi}$ (i.e., $g_+<g<g_-$), and $v_\pm =
\sqrt{1\pm g U_0/\pi}$. The inter-wire interaction also leads to
backscattering of amplitude $E_F^2 U_1/2 \pi ^2$ in the ``$-$''mode,
making evolution of this mode subject to the sin-Gordon Lagrangian.
For the zero chemical potential $\mu \equiv v_- \Delta k_F=0$, 
a renormalization-group analysis \cite{sol} of the uniform
sin-Gordon model at energies larger than $T_L$ leads to the
renormalized values of the parameters in (\ref{L}) for the $-$
mode. When $g_->1$, the backscattering amplitude goes to zero
if its initial value is sufficiently small, $g_- -1>|U_1| v_- /2\pi$.
Otherwise, the backscattering amplitude increases and becomes of
order $1$, while $g_-$ renormalizes to $1/2$ as the energy cut-off
scales down to the gap $M_0$ opening in the $-$ mode and evaluated as 
$M_0 \simeq E_F (|U_1| v_-/4 \pi^2(1-g_-))^{[2(1-g_-)]^{-1}}$ for 
$1-g_- \gg |U_1|v_-/4 \pi^2$. A finite $\mu$ does not change this behavior 
unless $\mu > M_0$. Since $U_0>|U_1|$
for realistic interactions, if $g=1$, this scaling always brings the
"-" mode into the gapless TLL regime. Therefore, existence of the
{\em interacting repulsive} TLL's in the individual wires with $g<1$
is a prerequisite for the opening of the gap in the energy spectrum of
the relative density excitations and associated strong Coulomb drag.
Assuming the gap, we construct below an effective low-energy model
for electron transport taking into account impurity scattering inside the
wires which introduces the term
\begin{equation}
{\cal L}_{imp}= - \frac{2 E_F}{\pi } \int_0^L dx \sum_j \left[
A_j(x) e^{i  \phi_j(x,t)} + h.c. \right],
\label{2} \end{equation}
to the total Lagrangian. Here $A_{1,2}(x)$ are the amplitudes of
weak one-electron backscattering in the two wires.

\emph{Duality Transformation.} An effective model for energies lower
than some cut-off $D'$ specified below can be derived from the
expression for the partition function ${\cal Z}$ associated with the
combined Lagrangians (\ref{L}) and (\ref{2}) following Schmid
\cite{schmid}. Without impurities, the $-$ and $+$ modes are
decoupled. Integrating out $\phi_-$ in the reservoirs, we see that the
``$-$'' part of ${\cal Z}$ describes rare tunneling of the massive $-$
mode between neighboring degenerate vacua characterized by
the quantized values of $\sqrt{2} \phi_-(\tau,x)+ 2 \mu x = 2 \pi m$,
where $m$ is integer. Variation of $m$ by $\pm 1$ corresponds to
tunneling of an instanton (anti-instanton) through the wire. The
tunneling amplitude has been found as $P e^{-M/T_L}$, with the
energy gap $M=\sqrt{ M_0^2-\mu^2} \equiv M_0 \sin \varpi$, from 
evaluation \cite{d3} of the instanton mass $M_0$ , 
and by mapping onto a free fermionic model \cite{1}. 
The instanton calculation \cite{2} also
gives the prefactor $P=C \times \sqrt{D'} (\sin^3\varpi M_0 T_L)^{1/4}$
up to a constant $C$ of order 1. A high-energy cut-off $D'$ of the
long-time asymptotics of the instanton-instanton interaction $F(\tau)=
\ln\{\sqrt{\tau^2 + 1/ D'^2}\}$ created by the reservoirs varies
with $\mu$ from $D' \simeq \sqrt{M_0 T_L}$ for $\mu=0$, to $D' \simeq
(M_0/\mu) T_L$ for $\mu > T_L$.

Weak impurity pinning does not change the form of instantons or
their dynamics as long as $E_F \int dx |A_{1,2}(x)| \ll M$. However,
it couples the $\pm$ modes, since the impurity Lagrangian (\ref{2})
restricted to the $m$th vacuum of the $-$ mode creates a scattering
potential for the $+$ mode
\begin{eqnarray}
\lefteqn{{\cal L}'_{imp}= - \frac{2 E_F (-1)^m}{\pi }} & &
\nonumber \\
& & \times \int_0^L dx \left[ e^{i  \phi_+(x,\tau)/\sqrt{2}}
\sum_j A_j(x) e^{i  (-1)^j \mu x} + h.c. \right] \, .
\label{2'} \end{eqnarray}
The sign $(-1)^m$ of this potential varies as vacuum is switched from
$m$ to $m \pm 1$ by instanton tunneling. For energies smaller than $T_L'
\equiv v_+/L$, the Lagrangian (\ref{2'}) is equivalent to that of an
effective point-like scatterer, \( (-1)^m (2 V_{imp}T_L'/
\pi ) \cos(\phi_+(\tau,0)/ \sqrt{2} )\), in the uniform
non-interacting ($g=1$) TLL, where the amplitude $V_{imp}$ is
proportional to $A_{1,2}$ in the lowest order \cite{1}:
$V_{imp}\simeq (E_F/T_L')^{1-g_+/2} | \int dx \left(\sum_j
A_j(x) e^{i(-1)^j \mu x} \right)|$. Variation of the
amplitude sign in Eq.\ (\ref{2'}) can be accounted for by
an auxiliary pseudospin variable with the corresponding Pauli
matix $\sigma_3$. The energy splitting becomes an operator
$\sigma_3 (2 V_{imp} T_L'/\pi) \cos(\phi_s(\tau,x_0)/\sqrt{2})$
acting on the pseudospin, and every (anti-)instanton tunneling
reverses the $\sigma_3$ values with the Pauli matrix
$\sigma_1$. The partition function can then be written as
%
\begin{eqnarray}
{\cal Z} &\propto& \sum_{N=0}^\infty \sum_{a_j=\pm}\int D \phi_s
{e^{-{\cal S}_0[\phi_+]} \over N!} \nonumber \\
& & \times Tr_\sigma [
T \left\{\int \left(\prod_{i=1}^N d \tau_i P e^{-{M \over T_L}}
\sigma_1(\tau_i)\right)
\right.  \label{3}  \\
& &\times \left.
e^{\sum_{i,j} {a_i a_j \over 2}  F(\tau_i-\tau_j) +
\frac{2 T_L' V_{imp}}{\pi }  \int d \tau \sigma_3(\tau)
\cos({\phi_+(\tau,0) \over \sqrt{2}})} \right\} ] \ .
\nonumber
\end{eqnarray}

\noindent
Here ${\cal S}_0[\phi_+]=\int_0^\beta d \tau {\cal L}_+$ with $g_+=1$
is the free TLL Eucleadian action, and $T$ denotes time-ordering. All
$\tau$-integrals run from $0$ to inverse temperature $\beta$, and
$\sum_j a_j=0$. To have all $\sigma_{1,3}$-matrices time-ordered in
(\ref{3}), we attributed
each of them to a corresponding time $\tau$ assuming that their time
evolution is trivial. Noticing that the interaction $F$ coincides with
the pair correlator of some bosonic field $\theta_-$ whose evolution
is described with the free TLL action ${\cal S}_0[\theta_-]$,
we rewrite (\ref{3}) ascribing the factors $\exp \{ \mp i \theta_-
(\tau_j,0)/\sqrt{2} \}$ to the $\tau_j$ instantons/anti-instantons:

\begin{eqnarray}
{\cal Z}&\propto& Tr_\sigma  \left\{ T  \int D \phi_+ D \theta_-
e^{- {\cal S}_0[\phi_+]-{\cal S}_0[\theta_-] + {\cal S}'[\sigma_{1,3},
\phi_+, \theta_-]}\right\} \label{4}\\
{\cal S}'&=&2 \int d \tau [
P e^{-{M \over T_L}} \sigma_1(\tau)\cos({\theta_-(\tau,0)\over
\sqrt{2}})) \nonumber \\
& & \ \ \ \ \ \ \  + \frac{ T_L' V_{imp}}{\pi } \sigma_3(\tau)
\cos({\phi_+(\tau,0) \over \sqrt{2}})] \ .
\nonumber
\end{eqnarray}
\noindent
The action in (\ref{4}) is similar to a model of point scatterer
with internal degree of freedom in the two-component TLL. This ``dual''
model (\ref{4}) is equivalent to the initial one (\ref{L}) at low
energies. Its imaginary time evolution is described in the operator
formalism as ${\cal Z} \propto Tr\{ e^{-\beta {\cal H}}\}$ with
a Hamiltonian ${\cal H}$ related to the action in (\ref{4}) in a
standard way. The model can be solved exactly through fermionization.

\emph{Fermionization.} From the commutation relations and
hermiticity, the Pauli matrices can be written as $\sigma_\alpha
= (-1)^{\alpha+1}{i\over 2} \sum_{\beta, \gamma} \epsilon^{\alpha,
\beta, \gamma} \xi_\beta \xi_\gamma$ with Majorana fermions
$\xi_{1,2,3}$ and antisymmetrical tensor $\epsilon$:
$\epsilon^{123}=1$. Point-like nature of the interaction in
(\ref{4}), and appropriate time dependence of the $\theta_-$ and
$\phi_+$ correlators, allow us to fermionize them introducing the
operators:
\[ \psi_-(0)=-\sqrt{D' \over 2\pi }\xi_3 e^{i{ \theta_-(0)
\over \sqrt{2}}}, \;\;
\psi_+(0)=\sqrt{T_L'\over 2 \pi}\xi_1 e^{i{\phi_+(0)
\over \sqrt{2}}}\, .
\]
Here $\psi_{\pm}(0)$ are the $x=0$ values of the fermionic fields
attributed to fluctuations of the total and relative densities.
These fields have linear dispersion relations ``inherited''
from the bosonic fields, with the cut-offs $T_L'$ and $D'$,
respectively. Substitution of these fields into the Hamiltonian of the
dual model (\ref{4}) makes it free-electron-like with the interaction 
reduced to
tunneling between the $\psi_{\pm}$ fermions and the Majorana
fermion $\xi_2 \equiv \xi$.

Application of voltages $V_{1,2}$ across the first and second
wires can be described \cite{PRB}
with a gauge transformation $\phi_{1,2} \to \phi_{1,2} -V_{1,2}t$
in the real-time Lagrangian (\ref{2}). It is equivalent to
transformation $\phi_{\pm} \to \phi_{1,2} -\sqrt{2} V_{\pm}t, \
V_\pm \equiv (V_1 \pm V_2)/2$, in the arguments of  the
$\cos$-terms in the Lagrangians (\ref{L}) and (\ref{4}). Since
each instanton tunneling in the $-$ mode transfers charge $\Delta
\phi_-/\sqrt{2} \pi=1$ and energy $-V_-$, this transformation of
$\phi_-$ causes a shift $\theta_-/  \sqrt{2} \rightarrow \theta_- /
\sqrt{2}+ V_- t$ of the $\cos$-argument in the real-time form of
the action (\ref{4}). Assuming that both voltages are sufficiently
small, $|V_{1,2}|<T_L<M$, we neglect their effect on the other
parameters. The real-time Lagrangian associated with the
fermionized Hamiltonian is:
%
%
\begin{eqnarray}
{\cal L}_{F}=i \xi \partial_t \xi(t) &+& i\sum_{\pm} \large\{
\int d\!x \psi^+_\pm(x,t) (\partial_t + \partial_x) \psi_\pm(x,t)
 \nonumber \\
&-&\sqrt{\Gamma_\pm } [\psi^+_\pm(0,t)\xi(t) e^{\pm iV_\pm t}+h.c.]
\large\} \ ,
\label{6}
\end{eqnarray}
%
\noindent
where the rate of impurity scattering is $\Gamma_+=2T_L' V_{imp}^2/
\pi$ and the rate of the instanton tunneling is $\Gamma_-=2 \pi C^2
\sqrt{T_L M_0 \sin^3 \varpi} e^{-2M/T_L}$. The currents flowing
through the first and second wire can be related as $J_{1,2}=
(J_+ \pm J_-)/\sqrt{2}$ to the currents of the $\pm$ modes
$J_\pm=- \partial_t \phi_\pm/2 \pi$. Duality Transformation
makes $J_- $ equal to the tunneling current 
$J_- = -i\sqrt{\Gamma_ -/2}[\psi^+_-(0,t)\xi(t)
e^{-iV_- t}-h.c.]$ in the fermionic model (\ref{6}). Meanwhile, the 
tunneling current of the $+$ mode $j_+ =i\sqrt{\Gamma_ +/2}[\psi^+_+(0,t)
\xi(t) e^{iV_+ t}-h.c.]$ coincides with the backscattered current of the 
bosonic Lagrangian (\ref{2}), 
whose average is simply related to the average  direct current \cite{PRB}:
$\langle J_+ \rangle =V_+/(\sqrt{2}\pi ) - \langle j_+\rangle$.

\emph{Currents}. Since the Lagrangian (\ref{6}) is Gaussian, both
currents can be found with the non-equilibrium Keldysh technique as
$\langle J_- \rangle =(\sqrt{2}\Gamma_-/ \pi )J(V_-/T,\Gamma/T)$ and
$\langle j_+
\rangle =(\sqrt{2}\Gamma_+/ \pi )J(V_+/T,\Gamma/T)$, where
\begin{equation}
J\left({V \over T}, {\Gamma \over T}\right) =  \Gamma \int d\omega
\frac{f({\omega-V \over T})-f({\omega +V \over T})}{\omega^2+4
\Gamma^2} \, .
\label{9} \end{equation}
In spite of the coupling between the $\pm$ modes, each of the
currents $\langle J_\pm \rangle$ is independent of the
voltage $V_\mp$ applied to the opposite mode. At $T=0$, eq.\ (\ref{9})
gives for the currents through the first and the second wire:
\[
\langle J_{1,2} \rangle ={V_+ \over 2 \pi} - {1 \over \pi }\left[
\Gamma_+ \arctan \left(\frac{V_+}{2\Gamma}\right) \mp \Gamma_-
\arctan \left(\frac{V_-}{2\Gamma}\right) \right] \, .
\]

The effect of the Coulomb drag is clearly seen when a voltage $V$ is
applied to the first (drive) wire only, $V_+=V_-=V/2$. In this case,
the average drag current in the second wire $\langle J_2 \rangle =
{V \over 4 \pi}- {\Gamma \over \pi }\arctan (V/4\Gamma)$ depends
only on the total crossover energy $\Gamma$. The current is suppressed
below this energy and approaches a half of the maximum total current
$\sigma_0 V$ above it, where $\sigma_0 \equiv e^2/2\pi \hbar$ is the
conductance of a ballistic 1D conductor. This shows that the
Fermi-liquid reservoirs
and one-electron impurity scattering have similar effect of
suppressing the drag current. However, both mechanisms affect
the drive wire current differently, $\langle J_1 \rangle =
{V \over 4 \pi}+ {\Gamma_- - \Gamma_+ \over \pi }\arctan
(V/4\Gamma)$. Below the crossover, this current reaches
maximum $\sigma_0 V$ in absence of disorder, while the disorder
suppresses it. Above the crossover, the current is about
a half of the maximum, and the conductance $\langle J_1 \rangle/V$
takes on the fractional value $\sigma_0/2$, which is not affected
by the disorder.

In the regime linear in $V_\pm$, the transport is characterised
by the ``diagonal'' conductance $G_C$ and trans-conductance $G_T$. To
find them we use that $J(V/T,\Gamma/T)=V \psi'(1/2+\Gamma/(\pi T))/(2
\pi T)\equiv G(T/\Gamma)V/(2\Gamma)$ for small $V$. Here
$\psi'(x)$ is the derivative of the di-gamma function, and the
function $G$ decreases monotonically from one at zero temperature,
$G=1-{1 \over 3}(\pi T/2 \Gamma)^2$, to zero at large
temperatures, $G=(\pi \Gamma/2T)-14 \zeta(3)(\pi \Gamma/2T)^2$, where
the zeta function $\zeta(3) \approx 1.2$. The linear trans-conductance
$G_T=\sigma_0 (1-G)/2$ is a function of $T/\Gamma$; it approaches 1/2
at large temperatures $T\gg \Gamma$ and vanishes at $T\ll \Gamma$.
The diagonal conductance $G_C= \sigma_0 \left[ 1/2 + G(T/\Gamma
)(\Gamma_- -\Gamma_+)/ \Gamma  \right]$ has a fractional
large-temperature asymptotics $\sigma_0 [1/2 +(\Gamma_- - \Gamma_+)
\pi /(4T)]$. The sign of $\Gamma_- - \Gamma_+$ determines
whether the conductance increases or decreases with decreasing
temperature. At zero temperature, $G_C= \sigma_0 \Gamma_-/\Gamma$ and
varies from 0 for $\Gamma_+ \gg \Gamma_-$ to its maximum $\sigma_0$
in absence of disorder.  
\begin{figure}[htb]
\begin{center}
\leavevmode
\psfig{file=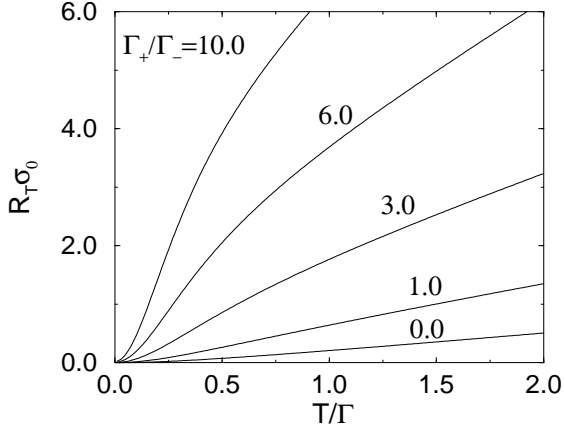,width=2.9in,angle=0}
\narrowtext{ \caption{
           Trans-resistance $R_T$ due to the current drag between
           two single-mode conductors as a fnction of temperature for
           different rates $\Gamma_+$ of impurity scattering in the
           conductors.}}
\end{center}
\end{figure}
In experiments, the drag trans-resistance $R_T$ is used typically
(see, e.g. \cite{debray,tarucha}) as the quantitative measure of
strength of the Coulomb drag. We obtain $R_T$ by inversion of the
conductance matrix, $R_T=G_T/(G_C^2-G_T^2)$, and show the results
in Fig. 1. Without impurity scattering ($\Gamma_+=0$),
$G_C+G_T=\sigma_0$ and the trans-resistance reduces to a one-parameter
function, $\bar{R}_T(T/\Gamma_-)=(1-G(T/\Gamma_-))/[2 \sigma_0 G]$.
It increases with temperature from zero at $T=0$ first quadratically,
$\bar{R}_T=\pi [{1 \over 3} (\pi T/2 \Gamma_-)^2-
{2 \over 15} (\pi T/2 \Gamma_-)^4]$ and then, above $\Gamma_-$, linearly:
$\bar{R}_T=2T/\Gamma_-+[56 \zeta(3)/\pi^2-\pi]$ up to an exponentially large
value $\bar{R}_T \propto \exp[M/T_L]$ at the upper boundary of
applicability of our model $T \approx T_L$. The impurity scattering
changes the trans-resistance to $R_T=\bar{R}_T(T/\Gamma){\Gamma \over
\Gamma_-} [1-{\Gamma_+ \over \Gamma}G(T/\Gamma)]^{-1}$. In particular,
$R_T$ is always enhanced by the scattering at temperatures larger
than $\Gamma$ as $R_T=\bar{R}_T(T/\Gamma_-)+{56 \over \pi^2}\zeta(3)
\Gamma_+/\Gamma_-$ and remains small below $\Gamma_-$ (see Fig.\ 1).

The results obtained in this work are valid for the 
following hierarchy of energies, $V,T\ll T_L \ll M$, and for sufficienly 
small amplitude of impurity scattering, $V_{imp}<M/T_L(T_L/E_F)^{g_+/2}$. 
They show that without impurity scattering, the linear trans-conductance 
$G_T$ is exponentially close to $\sigma_0/2$ at $\Gamma \ll T \le T_L$. 
$G_T$ should remain exponentially close to 
$\sigma_0/2$ even at larger temperature, $T \ge T_L$, but with 
deviations from this value now due to the thermal activation of 
instantons, $G_T-\sigma_0/2 \propto e^{-M/T}$. With further increase 
of temperature beyond $M$, $G_T$ is expected to deviate from 
$\sigma_0/2$, decreasing as $1/T$, as can be seen from a solution \cite{1} 
of a similar problem. When $T_L$ approaches $M$ (e.g., with decreasing 
conductor length) the width of the peak in the $G_T(T)$ 
dependence around $T_L$ reduces and this dependence smoothly goes over into 
the perturbative regime. In this case, similarly to the strong 
coupling regime, $G_T\propto T^2$ at low temperatures, $T\ll T_L$, 
but $G_T \propto T^{4g_- -3}$ at $T\gg T_L$. The trans-conductance 
can even grow with $T$, in particular, $G_T \propto T$ 
\cite{gurevich,rv,debray} for 1D Fermi liquid conductors \emph{above} $T_L$.
Note, however, that the Fermi liquid description of the drag problem is inconsistent, since it takes into account only the intrawire part of the Coulomb
interaction.

To evaluate the minimum length $L_*=v_/M_0$ necessary for observation 
of the strong Coulomb drag, we notice that in the optimum situation for 
observation of drag, the wire parameters: the width $w$, their 
separation $d$, and the distance $D$ from each wire to its screening 
gate, are related by the set of inequalities: $w<D<d \simeq 1/k_F$. 
Standard expressions for the electrostatics of inter-wire and intra-wire 
interactions show that these inequalities provide sufficiently small 
TLL interaction 
parameter $g$, and not-too-small backscattering amplitude $U_1$ of the 
interwire interaction, while keeping the upward renormalization of 
$g_-$ caused by the forward scattering $U_0$ to a minimum. 
We evaluate $L_*\simeq 3 \mu m$ \cite{dd3} for $w=10$nm, 
$D=20$nm, $d=40$nm, $k_F^{-1}=30$nm. These parameters are realistic for the 
present-day GaAs heterostructures and close to satisfying the inequalities.

In conclusion, we have shown that strong Coulomb drag occurs
between currents of \emph{repulsive} TLL in two 1D conductors at
temperatures and/or voltage differences above some crossover energy
$\Gamma$ and below the energy gap $M$ of the relative density fluctuations.
The drag is suppressed below $\Gamma$ and grows to the ideal one at
$T,V \to T_L$.

$^*$ On leave of absence from A.F.Ioffe Physical Technical
Institute, 194021, St. Petersburg, Russia.


\begin{references}
\bibitem{d1} A. Komnik and R. Egger, Phys.\ Rev.\ Lett. {\bf 80},
2881 (1998). Note that the boundary conditions for a 1D conductor
connected to the reservoirs used in this work can be reduced to
rescaling of the bias voltage which is inconsistent \cite{PRB} with
the physics of inhomogeneous TLL.
\bibitem{dd3} A smaller estimate of $L_*$ in \cite{d3} 
resulted from not taking into account the renormalization of $g_-$.
\end{references}
\end{document}